\documentclass{wqcd03}                 

\usepackage{mathptmx}                 
\confname{QCD@Work 2003 - International Workshop on QCD, Conversano, Italy, 
14--18 June 2003}

\usepackage{epsfig}

\usepackage{rotate}

\newcommand{\lsim}{
\mathrel{\hbox{\rlap{\hbox{\lower4pt\hbox{$\sim$}}}\hbox{$<$}}}}

\newcommand{\gsim}{
\mathrel{\hbox{\rlap{\hbox{\lower4pt\hbox{$\sim$}}}\hbox{$>$}}}}

\title{Theoretical Review of CP Violation}

\author{Robert Fleischer}

\address{Theory Division, CERN, CH-1211 Geneva 23, Switzerland}

\begin{document}


\begin{titlepage}

\begin{flushright}
CERN-TH/2003-263\\
hep-ph/0310313
\end{flushright}

\vspace{2.0truecm}
\begin{center}
\Large\bf Theoretical Review of CP Violation
\end{center}

\vspace{1.0cm}
\begin{center}
{\large Robert Fleischer}\\[0.1cm]
{\sl Theory Division, CERN, CH-1211 Geneva 23, Switzerland}
\end{center}

\vspace{1.8truecm}

\begin{center}
{\large {\bf Abstract}}
\end{center}

\vspace{0.3cm}

\begin{quotation}
\noindent
The focus of our considerations is the $B$-meson system, which will provide 
stringent tests of the Kobayashi--Maskawa mechanism of CP violation in this 
decade. After a classification of the main possible strategies to achieve 
this goal, we discuss the status of the $B$-factory benchmark modes in view 
of the current experimental data. We shall then turn to the ``El Dorado'' for 
$B$-decay studies at hadron colliders, the $B_s$-meson system, where
we will also address new, theoretically clean strategies to explore
CP violation.
\end{quotation}

\vspace{2.0truecm}

\begin{center} 
{\sl Invited talk at QCD@Work 2003\\
Conversano (Bari), Italy, 14--18 June 2003\\
To appear in the Proceedings}
\end{center}

\vfill
\noindent
CERN-TH/2003-263\\
October 2003

\end{titlepage}

\thispagestyle{empty}
\vbox{}
\newpage
 
\setcounter{page}{1}


\begin{abstract}
The focus of our considerations is the $B$-meson system, which will provide 
stringent tests of the Kobayashi--Maskawa mechanism of CP violation in this 
decade. After a classification of the main possible strategies to achieve 
this goal, we discuss the status of the $B$-factory benchmark modes in view 
of the current experimental data. We shall then turn to the ``El Dorado'' for 
$B$-decay studies at hadron colliders, the $B_s$-meson system, where
we will also address new, theoretically clean strategies to explore
CP violation.
\end{abstract}

\maketitle


%
%
%
\section{Setting the Stage}\label{sec:intro}
\subsection{Preliminaries}
The discovery of CP violation in 1964 through the observation of 
$K_{\rm L}\to\pi^+\pi^-$ decays came as a big surprise \cite{CP-discovery}. 
As is well known, this particular manifestation of CP violation, which is 
described by the famous parameter $\varepsilon_K$, originates from the fact 
that the mass eigenstate $K_{\rm L}$ is not a pure CP eigenstate with 
eigenvalue $-1$, but one that receives also a tiny admixture of the 
CP-even eigenstate. After tremendous efforts, also ``direct'' CP violation, 
i.e.\ CP-violating effects arising directly at the decay-amplitude level, 
could be established by the NA48 (CERN) and KTeV (FNAL) collaborations 
in 1999, through a measurement of a non-vanishing value of 
$\mbox{Re}(\varepsilon_K'/\varepsilon_K)$ \cite{eps-prime}. Unfortunately, 
this observable does not allow a stringent test of the Standard-Model 
description of CP violation, unless significant theoretical progress 
concerning the relevant hadronic matrix elements can be made (for a
detailed discussion, see \cite{buras-KAON}).

In this decade, the exploration of CP violation is governed by $B$ mesons,
which provide various tests of the Kobayashi--Maskawa mechanism \cite{KM}, 
allowing us to accommodate this phenomenon in the Standard Model (SM). 
Moreover, these studies offer also interesting insights into hadron 
dynamics. We will hence focus on the $B$-meson system in the following 
discussion; a considerably more detailed review can be found in
\cite{RF-Phys-Rep}. At the moment, the experimental stage is governed
by the asymmetric $e^+e^-$ $B$ factories operating at the $\Upsilon(4S)$ 
resonance, with their detectors BaBar (SLAC) and Belle (KEK). These 
experiments have already established CP violation in $B_d\to J/\psi K_{\rm S}$
decays in 2001, which represents the beginning of a new era in the exploration 
of CP violation, and many interesting strategies can now be confronted with 
the data \cite{giorgi}. In the near future, we expect also interesting
$B$-physics results from run II of the Tevatron, which will provide -- 
among other things -- first access to decays of $B_s$ mesons \cite{TEV-BOOK}. 
In the era of the LHC, these modes can then be fully exploited 
\cite{LHC-BOOK}, in particular at LHCb (CERN) and BTeV (FNAL).

\begin{figure}
\vspace{0.10in}
\centerline{
\epsfysize=5.3truecm
\epsffile{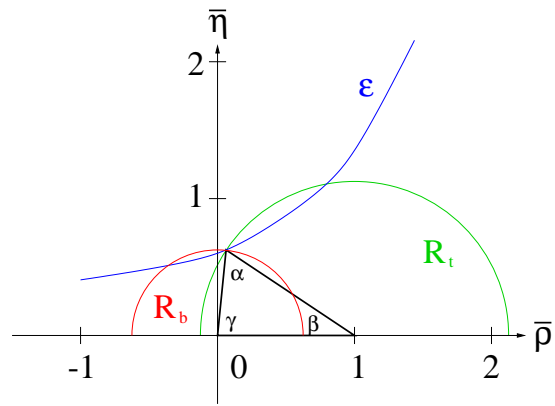}
}
\caption{Contours to determine the UT in the
$\overline{\rho}$--$\overline{\eta}$ plane.}\label{fig:cont-scheme}
\end{figure}

\subsection{Central Target: Unitarity Triangle}
The main goal is to overconstrain as much as possible the apex of the 
unitarity triangle (UT) of the Cabibbo--Kobayashi--Maskawa (CKM) matrix 
in the plane of the generalized Wolfenstein parameters $\overline{\rho}$ and 
$\overline{\eta}$ \cite{wolf}. On the one hand, we may obtain indirect
information on the UT angles through the ``CKM fits'' \cite{CKM-Book}, 
where the following ingredients enter, as illustrated in 
Fig.~\ref{fig:cont-scheme}: using semileptonic $B$ decays 
caused by $b\to u\ell\overline{\nu}_{\ell}, c\ell\overline{\nu}_{\ell}$ 
quark-level transitions, we may determine
the side $R_b\propto |V_{ub}/V_{cb}|$, whereas $R_t\propto |V_{td}/V_{cb}|$ 
can be determined, within the SM, with the help of 
$B^0_q$--$\overline{B^0_q}$ mixing ($q\in\{d,s\}$). Moreover, the SM
interpretation of $\varepsilon_K$ allows us to fix a hyperbola in the
$\overline{\rho}$--$\overline{\eta}$ plane. Following these lines, we
obtain the following typical ranges for the UT angles: 
\begin{equation}\label{UT-Fit-ranges}
70^\circ\lsim\alpha\lsim 130^\circ, \quad
20^\circ\lsim\beta\lsim 30^\circ, \quad
50^\circ\lsim\gamma\lsim 70^\circ.
\end{equation}
On the other hand, the measurement of CP-violating effects in $B$-meson 
decays allows us to obtain {\it direct} information on $\alpha$, $\beta$ 
and $\gamma$. In this context, non-leptonic transitions play the key r\^ole.
Within the SM, the unitarity of the CKM matrix allows us to write the
amplitude for any given non-leptonic $B$ decay in the following way:
\begin{eqnarray}
A(\overline{B}\to\overline{f})&=&e^{+i\varphi_1}
|A_1|e^{i\delta_1}+e^{+i\varphi_2}|A_2|e^{i\delta_2}\label{ampl-intro}\\
A(B\to f)&=&e^{-i\varphi_1}|A_1|e^{i\delta_1}+e^{-i\varphi_2}|A_2|
e^{i\delta_2},\label{ampl-intro-CP}
\end{eqnarray}
where the $\varphi_{1,2}$ are CP-violating weak phases, which are introduced
by the CKM matrix, while the CP-conserving strong amplitudes 
$|A_{1,2}|e^{i\delta_{1,2}}$ encode the hadron dynamics of the given decay,
i.e.\ QCD is at work in these quantities. Using these amplitude 
parametrizations yields
\begin{eqnarray}
\lefteqn{{\cal A}_{\mbox{{\scriptsize 
CP}}}^{\rm dir}\equiv\frac{\Gamma(B\to f)-
\Gamma(\overline{B}\to\overline{f})}{\Gamma(B\to f)+\Gamma(\overline{B}
\to \overline{f})}=}\label{CP-intro}\\
&&\frac{2|A_1||A_2|\sin(\delta_1-\delta_2)\,
\sin(\varphi_1-\varphi_2)}{|A_1|^2+2|A_1||A_2|\cos(\delta_1-\delta_2)\,
\cos(\varphi_1-\varphi_2)+|A_2|^2},\nonumber
\end{eqnarray}
which shows nicely that this kind of CP violation -- ``direct'' 
CP violation -- originates from different decay amplitudes with both different
weak and different strong phases. If such a CP asymmetry is measured, the
goal is to extract the weak phase difference $\varphi_1-\varphi_2$, as it
is related to the angles of the UT and is typically given by $\gamma$. 
However, we observe immediately that the strong amplitudes
\begin{equation}\label{ampl-struct}
|A|e^{i\delta}\sim\sum\limits_k
\underbrace{C_{k}(\mu)}_{\mbox{pert.\ QCD}} 
\times\,\,\, \underbrace{\langle\overline{f}|Q_{k}(\mu)
|\overline{B}\rangle}_{\mbox{``unknown''}}
\end{equation}
introduce large uncertainties into the game through non-perturbative 
hadronic matrix elements of local four-quark operators, which are poorly
known.

\subsection{Main Avenues to Explore CP Violation}
In order to tackle this challenging problem, we may follow three main 
avenues. First, we may try to calculate the hadronic matrix elements 
entering (\ref{ampl-struct}), which is a very challenging issue. 
Nevertheless, as was discussed at this workshop by C.T. Sachrajda and
Z. Wei, interesting progress could recently be made in this direction 
through the development of the QCD factorization and perturbative 
hard-scattering formalisms, as well as soft collinear effective theory 
(for a comprehensive review, see \cite{li-ref}).

In order to convert measurements of CP asymmetries of the kind specified in 
(\ref{CP-intro}) into solid information on the angles of the UT, 
it is desirable to reduce as much as possible the theoretical input on 
hadronic matrix elements. Such strategies, allowing in particular 
the determination of $\gamma$, are provided by fortunate cases, where 
we may eliminate the hadronic matrix elements through relations between 
different decay amplitudes: we distinguish between exact relations, 
involving pure tree-diagram-like decays of the kind 
$B\to DK$ or $B_c\to D D_s$, and relations, which follow from the 
flavour symmetries of strong interactions, involving $B_{(s)}\to \pi\pi, 
\pi K, KK$ decays (see \cite{RF-Phys-Rep} and references therein). 

The third avenue we may follow is to employ decays of neutral $B_q$ 
mesons ($q\in\{d,s\}$), where we may obtain interference effects between
$B^0_q$--$\overline{B^0_q}$ mixing and decay processes, leading to
another type of CP violation, ``mixing-induced'' CP violation, which allows
us to play many games. If such a transition is dominated by a single
weak amplitude, i.e.\ the sums in (\ref{ampl-intro}) and 
(\ref{ampl-intro-CP}) run only over a single term, the corresponding 
``unknown'' hadronic matrix element cancels in the mixing-induced CP 
asymmetry; the most important example in this context is the ``golden'' 
mode $B_d\to J/\psi K_{\rm S}$. It is also very interesting and useful 
to complement relations between different decay processes with such
CP-violating observables.

\section{The $B$-Factory Benchmark Modes}
\subsection{The Amplitude Relation Avenue: $B\to\pi K$}
These modes originate from $\overline{b}\to\overline{d}d\overline{s},
\overline{u}u\overline{s}$ quark-level processes, and may receive 
contributions both from penguin and from tree topologies, where the latter 
involve the UT angle $\gamma$. Since the ratio of tree to penguin
contributions is governed by the tiny CKM factor 
$|V_{us}V_{ub}^\ast/(V_{ts}V_{tb}^\ast)|\approx0.02$, $B\to\pi K$ decays 
are dominated by QCD penguins, despite their loop suppression. 
As far as electroweak (EW) penguins are concerned, their effects are 
expected to be negligible in the case of the $B^0_d\to\pi^-K^+$, 
$B^+\to\pi^+K^0$ system, as they contribute here only in colour-suppressed 
form. On the other hand, EW penguins may also contribute in colour-allowed 
form to $B^+\to\pi^0K^+$ and $B^0_d\to\pi^0K^0$, and are hence expected 
to be sizeable in these modes, i.e.\ of the same order of magnitude as 
the tree topologies.

\begin{figure}
\vspace*{-0.0cm}
$$\hspace*{-0.3cm}
\epsfysize=0.15\textheight
\epsfxsize=0.17\textheight
\epsffile{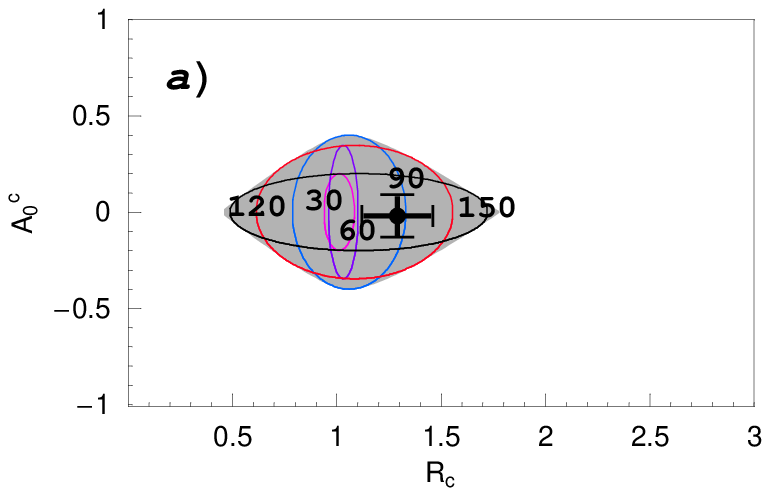} \hspace*{0.3cm}
\epsfysize=0.15\textheight
\epsfxsize=0.17\textheight
\epsffile{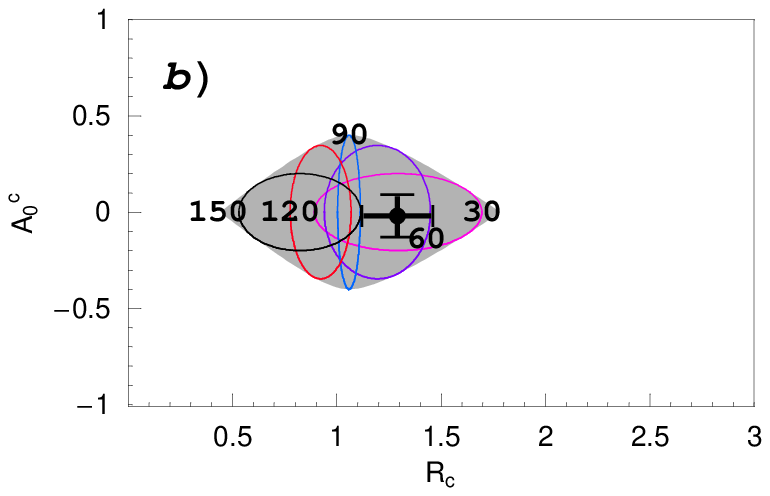}
$$
\vspace*{-0.6cm}
$$\hspace*{-0.3cm}
\epsfysize=0.15\textheight
\epsfxsize=0.17\textheight
\epsffile{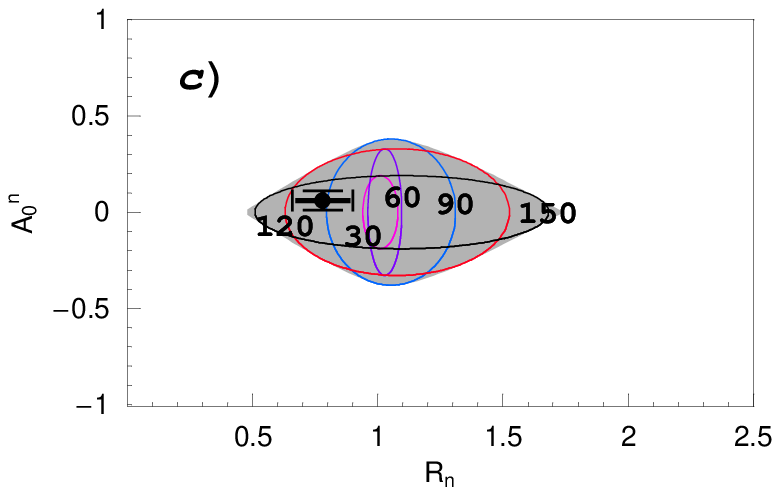} \hspace*{0.3cm}
\epsfysize=0.15\textheight
\epsfxsize=0.17\textheight
\epsffile{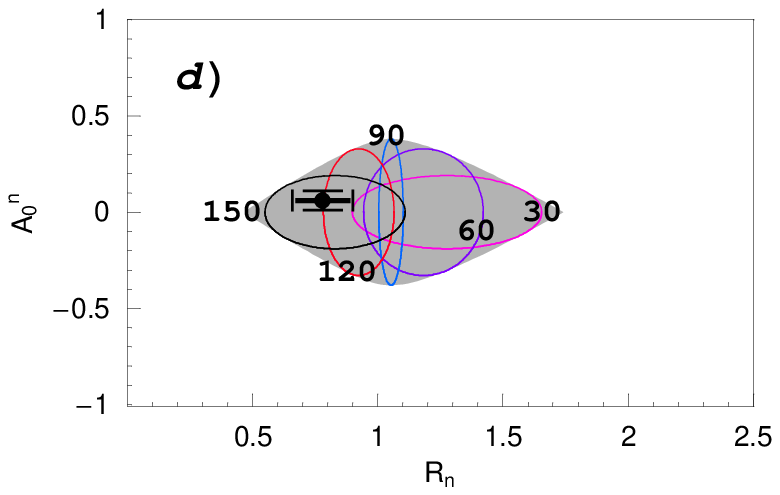}
$$
\vspace*{-0.8truecm}
\caption{The allowed regions in observable space of the charged
($r_{\rm c}=0.20$; (a), (b)) and neutral ($r_{\rm n}=0.19$; (c), (d))
$B\to \pi K$ systems for $q=0.68$: in (a) and (c), we show also the 
contours for fixed values of $\gamma$, whereas we give the curves 
arising for fixed values of $|\delta_{\rm c}|$ and $|\delta_{\rm n}|$ 
in (b) and (d), respectively.}\label{fig:BpiK-OS}
\end{figure}

Thanks to interference effects between tree and penguin amplitudes, 
we obtain a sensitivity on $\gamma$. In order to determine this angle, 
we may use an isospin relation as a starting point, suggesting the 
following combinations: the ``mixed'' $B^\pm\to\pi^\pm K$, 
$B_d\to\pi^\mp K^\pm$ system \cite{PAPIII}--\cite{defan}, the ``charged'' 
$B^\pm\to\pi^\pm K$, $B^\pm\to\pi^0K^\pm$ system 
\cite{NR}--\cite{BF-neutral1}, and the ``neutral'' $B_d\to\pi^0 K$, 
$B_d\to\pi^\mp K^\pm$ system \cite{BF-neutral1,BF-neutral2}. As noted in 
\cite{BF-neutral1}, all three $B\to\pi K$ systems can be described 
by the same set of formulae, just making straightforward replacements of 
variables. Let us first focus on the charged and neutral $B\to\pi K$ systems. 
In order to determine $\gamma$ and the corresponding strong phases, we have 
to introduce appropriate CP-conserving and CP-violating observables:
\begin{equation}\label{charged-obs}
\left.\begin{array}{l}R_{\rm c}\\A_0^{\rm c}\end{array}\right.
\equiv2\left[\frac{\mbox{BR}(B^+\to\pi^0K^+)\pm
\mbox{BR}(B^-\to\pi^0K^-)}{\mbox{BR}(B^+\to\pi^+K^0)+
\mbox{BR}(B^-\to\pi^-\overline{K^0})}\right]
\end{equation}
\begin{equation}\label{neutral-obs}
\left.\begin{array}{l}R_{\rm n}\\A_0^{\rm n}\end{array}\right.
\equiv\frac{1}{2}\left[\frac{\mbox{BR}(B^0_d\to\pi^-K^+)\pm
\mbox{BR}(\overline{B^0_d}\to\pi^+K^-)}{\mbox{BR}(B^0_d\to\pi^0K^0)+
\mbox{BR}(\overline{B^0_d}\to\pi^0\overline{K^0})}\right],
\end{equation}
where the $R_{\rm c,n}$ and $A_0^{\rm c,n}$ refer to the plus and 
minus signs, respectively. For the parametrization of these observables, 
we employ the isospin relation mentioned above, and assume that certain 
rescattering effects are small, which is in accordance with the 
QCD factorization picture \cite{QCD-fact}; large rescattering processes 
would be 
indicated by $B\to KK$ modes, which are already strongly constrained 
by the $B$ factories, and could be included through more elaborate 
strategies \cite{defan,neubert-BpiK,BF-neutral1}. Following these lines, 
we may write
\begin{equation}
R_{\rm c,n}=\mbox{fct}(q,r_{\rm c,n},\delta_{\rm c,n},\gamma), \quad
A_0^{\rm c,n}=\mbox{fct}(r_{\rm c,n},\delta_{\rm c,n},\gamma),
\end{equation}
where the parameters $q$, $r_{\rm c,n}$ and $\delta_{\rm c,n}$ have 
the following meaning: $q$ describes the ratio of EW penguin to tree 
contributions, and can be determined with the help of $SU(3)$ 
flavour-symmetry arguments, yielding $q\sim 0.7$ \cite{NR}. On the
other hand, $r_{\rm c,n}$ measures the ratio of tree to QCD penguin 
topologies, and can be fixed through $SU(3)$ arguments and data on 
$B^\pm\to\pi^\pm\pi^0$ modes \cite{GRL}, which give $r_{\rm c,n}\sim0.2$. 
Finally, $\delta_{\rm c,n}$ is the CP-conserving strong phase between the 
tree and QCD penguin amplitudes. Since we may fix $q$ and $r_{\rm c,n}$, 
the observables $R_{\rm c,n}$ and $A_0^{\rm c,n}$ actually depend only 
on the two ``unknown'' parameters $\delta_{\rm c,n}$ and $\gamma$. If we 
vary them within their allowed ranges, i.e.\ 
$-180^\circ\leq \delta_{\rm c,n}\leq+180^\circ$ and 
$0^\circ\leq \gamma \leq180^\circ$, we obtain an allowed region in the 
$R_{\rm c,n}$--$A_0^{\rm c,n}$ plane \cite{FlMa1,FlMa2}. Should the
measured values of $R_{\rm c,n}$ and $A_0^{\rm c,n}$ fall outside this 
region, we would have an immediate signal for new physics (NP). On the 
other hand, should the measurements lie inside the allowed range, 
$\gamma$ and $\delta_{\rm c,n}$ could be extracted. The value of $\gamma$ 
thus obtained could then be compared with the results of other
strategies, whereas the strong phase $\delta_{\rm c,n}$ would offer
interesting insights into hadron dynamics. This exercise can be performed 
separately for the charged and neutral $B\to\pi K$ systems. 

In Fig.~\ref{fig:BpiK-OS}, we show the allowed regions in the 
$R_{\rm c,n}$--$A_0^{\rm c,n}$ planes \cite{FlMa2}, where the 
crosses represent the averages of the current $B$-factory 
data. As can be read off from the contours in these figures, 
both the charged and the neutral $B\to \pi K$ data favour 
$\gamma\gsim90^\circ$, which would be in conflict with the $\gamma$ 
range in (\ref{UT-Fit-ranges}) following from the ``standard analysis'' 
of the UT.
Interestingly, the charged modes point towards $|\delta_{\rm c}|\lsim90^\circ$ 
(factorization predicts $\delta_{\rm c}$ to be close to 
$0^\circ$ \cite{Be-Ne}), whereas the neutral decays seem to prefer 
$|\delta_{\rm n}|\gsim90^\circ$. Since we do not expect $\delta_{\rm c}$ 
to differ significantly from $\delta_{\rm n}$, we
arrive at a ``puzzling'' picture of the kind that was already considered
a couple of years ago in \cite{BF-neutral2}. On the other hand, the
data for the mixed $B\to\pi K$ system fall well into the SM region in 
observable space and do not indicate any ``anomalous'' behaviour.
A detailed discussion of this ``$B\to\pi K$ puzzle'', which may be a 
manifestation of new physics in the EW penguin sector, and its relation 
to rare $B$ and $K$ decays, was recently given in \cite{BFRS}. It will
be very exciting to follow the evolution of the data.

\subsection{The Neutral $B$-Decay Avenue}
\subsubsection{Time-Dependent CP Asymmetries}
A particularly simple but very important special case arises for neutral 
$B_q$-meson decays ($q\in\{d,s\}$) into final CP eigenstates $|f\rangle$, 
which satisfy $({\cal CP})|f\rangle=\pm|f\rangle$. Here we obtain the 
following expression \cite{RF-Phys-Rep}:
\begin{eqnarray}
\lefteqn{\frac{\Gamma(B^0_q(t)\to f)-
\Gamma(\overline{B^0_q}(t)\to \overline{f})}{\Gamma(B^0_q(t)\to f)+
\Gamma(\overline{B^0_q}(t)\to \overline{f})}}\nonumber\\
&&=\left[\frac{{\cal A}_{\rm CP}^{\rm dir}
\cos(\Delta M_q t)+{\cal A}_{\rm CP}^{\rm mix}
\sin(\Delta M_q t)}{\cosh(\Delta\Gamma_qt/2)-
{\cal A}_{\rm \Delta\Gamma}\sinh(\Delta\Gamma_qt/2)}
\right],\label{fleischer-CP-asym}
\end{eqnarray}
where
\begin{equation}\label{fleischer-obs}
{\cal A}^{\mbox{{\scriptsize dir}}}_{\mbox{{\scriptsize CP}}}
\equiv\frac{1-\bigl|\xi_f^{(q)}\bigr|^2}{1+
\bigl|\xi_f^{(q)}\bigr|^2} \quad\mbox{and}\quad
{\cal A}^{\mbox{{\scriptsize mix}}}_{\mbox{{\scriptsize
CP}}}\equiv\frac{2\,\mbox{Im}\,\xi^{(q)}_f}{1+
\bigl|\xi^{(q)}_f\bigr|^2},
\end{equation}
with
\begin{equation}
\xi_f^{(q)}=\mp e^{-i\phi_q}
\left[\frac{A(\overline{B_q^0}\to\overline{f})}{A(B_q^0\to f)}\right],
\end{equation}
describe the ``direct'' and ``mixing-induced'' CP-violating observables,
respectively. In the SM, the CP-violating weak $B^0_q$--$\overline{B^0_q}$ 
mixing phase $\phi_q$ is associated with the well-known box diagrams, and 
is given by
\begin{equation}\label{fleischer-phiq-def}
\phi_q=2\,\mbox{arg}(V_{tq}^\ast V_{tb})
=\left\{\begin{array}{cc}
+2\beta & \mbox{($q=d$)}\\
-2\lambda^2\eta & 
\mbox{($q=s$),}
\end{array}\right.
\end{equation}
where $\beta$ is the usual angle of the UT. Looking at 
(\ref{fleischer-CP-asym}), we observe that $\Delta\Gamma_q$ provides 
another observable ${\cal A}_{\rm \Delta\Gamma}$, which is, however, not 
independent from those in (\ref{fleischer-obs}).

\subsubsection{$B_d\to J/\psi K_{\rm S}$}
One of the most famous $B$-meson decays, the ``golden'' mode 
$B_d^0\to J/\psi K_{\rm S}$ to extract $\sin2\beta$ \cite{bisa},
originates from $\overline{b}\to\overline{c}c\overline{s}$ quark-level 
processes. Within the SM, it receives contributions both from tree 
and from penguin topologies, so that we may write the decay amplitude
as follows:
\begin{equation}\label{BdpsiK-ampl2}
A(B_d^0\to J/\psi K_{\rm S})\propto\left[1+\lambda^2 a e^{i\theta}
e^{i\gamma}\right],
\end{equation}
where the hadronic parameter $a e^{i\theta}$ is a measure of the ratio of 
the penguin to tree contributions \cite{RF-BdsPsiK}. Since this quantity
enters in a doubly Cabibbo-suppressed way, and is na\"\i vely expected to 
be of ${\cal O}(\overline{\lambda})$, where $\overline{\lambda}=
{\cal O}(\lambda)={\cal O}(0.2)$ is a ``generic'' expansion 
parameter \cite{FM-BpsiK}, we arrive at  
\begin{equation}\label{BpsiK-CP-dir}
{\cal A}_{\rm CP}^{\rm dir}(B_d\to J/\psi K_{\rm S})=0+
{\cal O}(\overline{\lambda}^3)
\end{equation}
\begin{equation}\label{BpsiK-CP-mix}
{\cal A}_{\rm CP}^{\rm mix}(B_d\to J/\psi K_{\rm S})=-\sin\phi_d+
{\cal O}(\overline{\lambda}^3).
\end{equation}
The decay $B_d^0\to J/\psi K_{\rm S}$ and similar channels led to the
observation of CP violation in the $B$ system in 2001 \cite{giorgi}; 
the current status of $\sin\phi_d\stackrel{\rm SM}{=}\sin2\beta$ is given 
as follows:
\begin{equation}
\sin2\beta=\left\{\begin{array}{ll}
0.741\pm 0.067  \pm0.033  &
\mbox{(BaBar \cite{Babar-s2b-02})}\\
0.733\pm 0.057  \pm0.028  &
\mbox{(Belle \cite{Belle-s2b-02}),}
\end{array}\right.
\end{equation}
yielding the world average 
\begin{equation}\label{s2b-average}
\sin2\beta=0.736\pm0.049, 
\end{equation}
which agrees well with the results of the ``CKM fits'' of the UT summarized
in (\ref{UT-Fit-ranges}), implying $0.6\lsim\sin2\beta\lsim0.9$. 

In the LHC era \cite{LHC-BOOK}, the experimental accuracy of
$\sin2\beta$ may reach a level requiring deeper insights into the 
corrections affecting (\ref{BpsiK-CP-mix}). A possibility to control them 
is provided by the $B_s\to J/\psi K_{\rm S}$ channel \cite{RF-BdsPsiK}. 
Moreover, also direct CP violation in $B_d\to J/\psi K_{\rm S}$ allows 
us to probe these effects. So far, there are no experimental indications 
for a non-vanishing value of 
${\cal A}_{\rm CP}^{\rm dir}(B_d\to J/\psi K_{\rm S})$.

The agreement between (\ref{s2b-average}) and the ``CKM fits'' is striking. 
However, it should not be forgotten that NP may nevertheless hide in 
${\cal A}_{\rm CP}^{\rm mix}(B_d\to J/\psi K_{\rm S})$. The point is that 
the key quantity is actually $\phi_d$ itself, which is given by
\begin{equation}\label{phid-det}
\phi_d=\left(47\pm4\right)^\circ \, \lor \,
\left(133\pm4\right)^\circ.
\end{equation}
Here the former value agrees perfectly with
$40^\circ\lsim2\beta\stackrel{\rm SM}{=}\phi_d\lsim60^\circ$, which is implied 
by the ``CKM fits'', whereas the latter would correspond to NP. The two 
solutions can obviously be distinguished through a measurement of the sign 
of $\cos\phi_d$. To accomplish this important task, several strategies were
proposed \cite{ambig}, but their practical implementations are unfortunately
rather challenging. One of the most accessible approaches employs 
the time-dependent angular distribution of the 
$B_d\to J/\psi[\to\ell^+\ell^-] K^\ast[\to\pi^0K_{\rm S}]$ decay products, 
allowing us to extract $\mbox{sgn}(\cos\phi_d)$ if we fix the sign of a 
hadronic parameter $\cos\delta_f$, which involves a strong phase $\delta_f$, 
through factorization \cite{DDF2,DFN}.

\subsubsection{$B_d\to\phi K_{\rm S}$}
Another important testing ground for the SM description of CP violation
is provided by the decay $B_d\to\phi K_{\rm S}$, which originates from
$\overline{b}\to\overline{s}s\overline{s}$ quark-level processes. In
analogy to its charged counterpart $B^\pm \to \phi K^\pm$, this mode is 
governed by QCD penguins \cite{BphiK-old}, but also EW penguin contributions 
are sizeable \cite{RF-EWP,DH-PhiK}. Since such penguin topologies are 
absent at the tree level in the SM, $B\to\phi K$ decays represent a 
sensitive probe for NP effects. Within the SM, we obtain the 
following relations \cite{growo}--\cite{GLNQ}:
\begin{equation}
{\cal A}_{\rm CP}^{\rm dir}(B_d\to \phi K_{\rm S})=0+
{\cal O}(\overline{\lambda}^2)
\end{equation}
\begin{equation}
{\cal A}_{\rm CP}^{\rm mix}(B_d\to \phi K_{\rm S})=
{\cal A}_{\rm CP}^{\rm mix}(B_d\to J/\psi K_{\rm S})+
{\cal O}(\overline{\lambda}^2).
\end{equation}

The current experimental status of the CP-violating $B_d\to\phi K_{\rm S}$
observables is given as follows \cite{Browder,Belle-BphiK}:
\begin{equation}\label{aCP-Bd-phiK-dir}
{\cal A}_{\rm CP}^{\rm dir}
=
\left\{\begin{array}{ll}
-0.38\pm0.37\pm0.12 &\mbox{(BaBar)}\\
+0.15\pm0.29\pm0.07 &\mbox{(Belle)}
\end{array}\right.
\end{equation}
\begin{equation}
{\cal A}_{\rm CP}^{\rm mix}
=
\left\{\begin{array}{ll}
-0.45\pm0.43\pm0.07 &\mbox{(BaBar)}\\
+0.96\pm0.50^{+0.11}_{-0.09} &\mbox{(Belle).}
\end{array}\right.
\end{equation}
Since we have, on the other hand, 
${\cal A}_{\rm CP}^{\rm mix}(B_d\to J/\psi K_{\rm S})=-0.736 \pm 0.049$, 
we arrive at a puzzling situation, which has already stimulated many
speculations about NP effects in $B_d\to\phi K_{\rm S}$ \cite{BPhiK-NP}.
However, because of the large experimental uncertainties and the
unsatisfactory current situation, it seems too early to get too excited by 
the possibility of having large NP contributions to the 
$B_d\to\phi K_{\rm S}$ decay amplitude. It will be very interesting to 
observe how the $B$-factory data will evolve, and to keep an eye on 
$B_d\to \eta' K_{\rm S}$ and other related modes.

\subsubsection{$B_d\to\pi^+\pi^-$}\label{ssec:Bpipi}
The $B_d^0\to\pi^+\pi^-$ channel is another prominent $B$-meson
transition, originating from $\overline{b}\to\overline{u}u\overline{d}$ 
quark-level processes. In the SM, we may write
\begin{equation}\label{Bpipi-ampl}
A(B_d^0\to\pi^+\pi^-)\propto\left[e^{i\gamma}-de^{i\theta}\right],
\end{equation}
where the CP-conserving strong parameter $d e^{i\theta}$ measures the 
ratio of the penguin to tree contributions \cite{RF-BsKK}. In contrast to 
the $B_d^0\to J/\psi K_{\rm S}$ amplitude (\ref{BdpsiK-ampl2}), this 
parameter does {\it not} enter (\ref{Bpipi-ampl}) in a doubly 
Cabibbo-suppressed way, thereby leading to the well-known ``penguin problem'' 
in $B_d\to\pi^+\pi^-$. If we had negligible penguin contributions, i.e.\ 
$d=0$, things would simplify as follows:
\begin{eqnarray}
{\cal A}_{\rm CP}^{\rm dir}(B_d\to\pi^+\pi^-)&=&0\\
{\cal A}_{\rm CP}^{\rm mix}(B_d\to\pi^+\pi^-)&=&\sin(\phi_d+2\gamma)
\stackrel{\rm SM}{=}-\sin 2\alpha,
\end{eqnarray}
where we have used $\phi_d\stackrel{\rm SM}{=}2\beta$ and the unitarity 
relation $2\beta+2\gamma=2\pi-2\alpha$ in the last identity. We observe 
that actually $\phi_d$ and $\gamma$ enter directly 
${\cal A}_{\rm CP}^{\rm mix}(B_d\to\pi^+\pi^-)$, and not $\alpha$. 
Consequently, since $\phi_d$ can straightforwardly be fixed through 
$B_d\to J/\psi K_{\rm S}$, we may use $B_d\to\pi^+\pi^-$ to probe 
$\gamma$. The current status of the CP-violating $B_d\to\pi^+\pi^-$ 
observables is given as follows:
\begin{equation}\label{Adir-exp}
{\cal A}_{\rm CP}^{\rm dir}
=\left\{
\begin{array}{ll}
-0.19\pm0.19\pm0.05 & \mbox{(BaBar \cite{BaBar-Bpipi})}\\
-0.77\pm0.27\pm0.08 & \mbox{(Belle \cite{Belle-Bpipi})}
\end{array}
\right.
\end{equation}
\begin{equation}\label{Amix-exp}
{\cal A}_{\rm CP}^{\rm mix}
=\left\{
\begin{array}{ll}
+0.40\pm0.22\pm0.03& \mbox{(BaBar \cite{BaBar-Bpipi})}\\
+1.23\pm0.41 ^{+0.07}_{-0.08} & \mbox{(Belle \cite{Belle-Bpipi}).}
\end{array}
\right.
\end{equation}
The BaBar and Belle results are not fully consistent with each other. If we 
calculate, nevertheless, the weighted averages of (\ref{Adir-exp}) and 
(\ref{Amix-exp}), we obtain 
\begin{eqnarray}
{\cal A}_{\rm CP}^{\rm dir}(B_d\to\pi^+\pi^-)&=&-0.39\pm0.16 \,\, (0.27)
\label{Bpipi-CP-averages}\\
{\cal A}_{\rm CP}^{\rm mix}(B_d\to\pi^+\pi^-)&=&+0.58\pm0.19 \,\, (0.34),
\label{Bpipi-CP-averages2}
\end{eqnarray}
where the errors in brackets are those increased by the PDG scaling-factor 
procedure \cite{PDG}. Should large direct CP violation in $B_d\to\pi^+\pi^-$,
as suggested by (\ref{Bpipi-CP-averages}), be confirmed by future data, we 
would require large penguin contributions with large CP-conserving strong 
phases. A significant impact of penguins on $B_d\to\pi^+\pi^-$ is also 
indicated by the data on the $B\to\pi K,\pi\pi$ branching ratios 
\cite{FlMa2,RF-Bpipi}, as well as by theoretical considerations 
\cite{Be-Ne,PQCD-appl,KMM}. Consequently, it is already evident that we 
{\it must} take the penguin contributions to $B_d\to\pi^+\pi^-$ into account.

\begin{figure}[t]
$$\hspace*{-1.cm}
\epsfysize=0.2\textheight
\epsfxsize=0.3\textheight
\epsffile{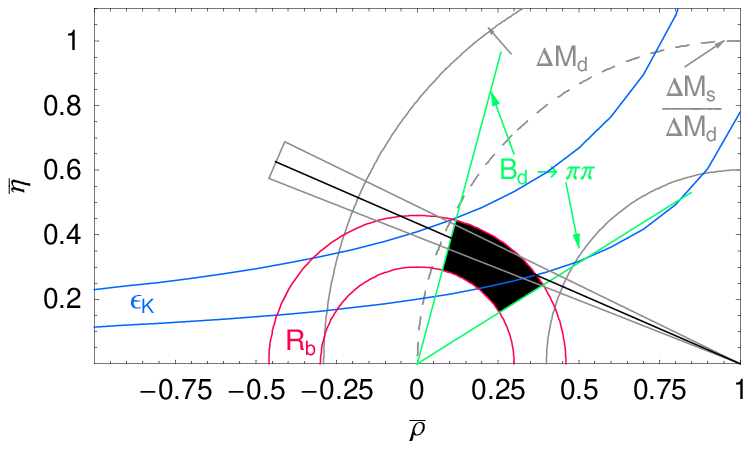} 
$$
\vspace*{0.0cm}
$$\hspace*{-1.cm}
\epsfysize=0.2\textheight
\epsfxsize=0.3\textheight
\epsffile{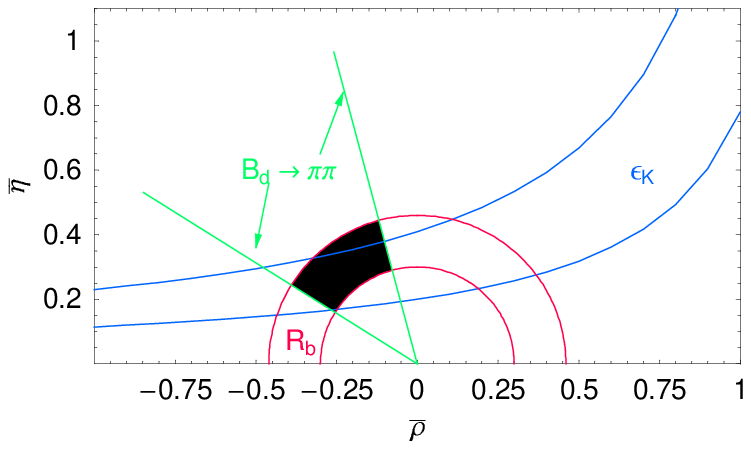}
$$
\caption{The allowed regions for the UT fixed through $R_b$ and CP 
violation in $B_d\to\pi^+\pi^-$, as described in the text: the upper
and lower figures correspond to $\phi_d=47^\circ$ and $\phi_d=133^\circ$, 
respectively ($H=7.5$).}\label{fig:rho-eta-Bpipi}
\end{figure}

A possibility to deal with this problem is provided by the strategies
proposed in \cite{FlMa2,RF-Bpipi}, employing $B_d\to\pi^\mp K^\pm$ (for 
alternative approaches, see \cite{RF-Phys-Rep}). If we make use of $SU(3)$
flavour-symmetry arguments and plausible dynamical assumptions, 
we may complement 
\begin{eqnarray}
{\cal A}_{\rm CP}^{\rm dir}(B_d\to\pi^+\pi^-)&=&
\mbox{fct}(d,\theta,\gamma)\label{ADIR-Bpipi}\\
{\cal A}_{\rm CP}^{\rm mix}(B_d\to\pi^+\pi^-)&=&
\mbox{fct}(d,\theta,\gamma,\phi_d)\label{AMIX-Bpipi}
\end{eqnarray}
with
\begin{equation}\label{H-def}
H=\left(\frac{1-\lambda^2}{\lambda^2}\right)
\left[\frac{{\rm BR}(B_d\to\pi^+\pi^-)}{{\rm BR}(B_d
\to \pi^\mp K^\pm)}\right]=\mbox{fct}(d,\theta,\gamma),
\end{equation}
allowing the extraction of $d$, $\theta$ and $\gamma$. Taking into 
account the $B$-factory result $H\sim7.5$, the CP asymmetries in 
(\ref{Bpipi-CP-averages}) and (\ref{Bpipi-CP-averages2}) point towards 
the following picture: for $\phi_d\sim47^\circ$, we obtain 
$\gamma\sim60^\circ$, in full accordance with the SM. On the other hand, 
the unconventional $\phi_d\sim133^\circ$ solution, which would require
CP-violating NP contributions to $B^0_d$--$\overline{B^0_d}$ mixing, 
favours $\gamma\sim120^\circ$. If we assume a scenario for physics
beyond the SM, where we have large NP contributions to 
$B^0_d$--$\overline{B^0_d}$ mixing, but not to the $\Delta B=1$ and
$\Delta S=1$ decay processes, which was already considered several years 
ago \cite{GNW} and can be motivated by generic arguments and within 
supersymmetry \cite{FIM}, we may complement $R_b$ (determined from
semileptonic tree decays) with the range for $\gamma$ extracted from our 
$B_d\to\pi^+\pi^-$ analysis, allowing us to fix the apex of the UT in 
the $\overline{\rho}$--$\overline{\eta}$ plane. The results of this exercise 
are summarized in Fig.~\ref{fig:rho-eta-Bpipi}, following \cite{FIM}, 
where also ranges for $\alpha$, $\beta$ and $\gamma$ are given and 
a detailed discussion of the theoretical uncertainties can be found. 
Interestingly, the measured branching ratio for the 
rare kaon decay $K^+\to\pi^+\nu\overline{\nu}$ seems to point towards 
$\gamma>90^\circ$ \cite{dAI}, thereby favouring the unconventional solution 
of $\phi_d=133^\circ$ \cite{FIM}. Further valuable information on this 
exciting possibility can be obtained from the rare decays 
$B_{s,d}\to\mu^+\mu^-$. We shall return to this issue in Subsection
\ref{ssec:BsKK}, discussing the $B_d\to\pi^+\pi^-$, $B_s\to K^+K^-$
system.

\section{The ``El Dorado'' for Hadron Colliders}
At the $e^+e^-$ $B$ factories operating at $\Upsilon(4S)$, $B_s$ mesons 
are not accessible. On the other hand, plenty of $B_s$ mesons will be 
produced at hadron colliders. Consequently, these particles are the 
``El Dorado'' for $B$-decay studies at run II of the Tevatron 
\cite{TEV-BOOK}, and later on at the LHC \cite{LHC-BOOK}. 

An important aspect of $B_s$ physics is the mass difference $\Delta M_s$
of the $B_s$ mass eigenstates, which can be complemented with $\Delta M_d$ to 
determine the side $R_t\propto|V_{td}/V_{cb}|$ of the UT. To this end, we 
use that $|V_{cb}|=|V_{ts}|$ to a good accuracy in the SM, and require 
just a single $SU(3)$-breaking parameter, which can be determined, 
e.g.\ on the lattice. At the moment, only experimental lower bounds on 
$\Delta M_s$ are available, which can be converted into upper bounds on 
$R_t$, implying $\gamma\lsim 90^\circ$ \cite{CKM-Book}. 
Once $\Delta M_s$ is measured, more stringent constraints on $\gamma$ 
will emerge. 

Another interesting quantity is $\Delta\Gamma_s$. While 
$\Delta\Gamma_d/\Gamma_d$ is negligibly small, $\Delta\Gamma_s/\Gamma_s$ 
may be as large as ${\cal O}(10\%)$ (for a recent study, see \cite{DGam}), 
thereby allowing interesting CP studies with ``untagged'' $B_s$ decay 
rates, where we do not distinguish between initially present $B^0_s$ 
or $\overline{B^0_s}$ mesons \cite{Bs-untagged}.

\subsection{$B_s\to J/\psi\phi$}\label{fleischer:sec-BsPsiPhi}
This promising channel is the $B_s$-meson counterpart of the ``golden''
mode $B_d\to J/\psi K_{\rm S}$, and is described by a transition amplitude 
with a completely analogous structure. In contrast to 
$B_d\to J/\psi K_{\rm S}$, the final state of $B_s\to J/\psi\phi$ 
is an admixture of different CP eigenstates, which can, however, be 
disentangled through an angular analysis of the 
$J/\psi [\to\ell^+\ell^-]\phi [\to\ K^+K^-]$ decay products \cite{DDLR,DDF}. 
Their angular distribution exhibits tiny direct CP violation, whereas 
mixing-induced CP-violating effects allow the extraction of
\begin{equation}\label{sinphis}
\sin\phi_s+{\cal O}(\overline{\lambda}^3)=\sin\phi_s+{\cal O}(10^{-3}).
\end{equation}
Since we have $\phi_s=-2\lambda^2\eta={\cal O}(10^{-2})$ in the SM, the 
determination of this phase from (\ref{sinphis}) is affected by
generic hadronic uncertainties of ${\cal O}(10\%)$, which may become an
important issue for the LHC era. These uncertainties can be controlled with
the help of flavour-symmetry arguments through the decay 
$B_d\to J/\psi \rho^0$ \cite{RF-ang}. Needless to note, the big hope is 
that experiments will find a {\it sizeable} value of $\sin\phi_s$, which 
would immediately signal the presence of NP contributions to 
$B^0_s$--$\overline{B^0_s}$ mixing.

Other interesting aspects of the $B_s\to J/\psi\phi$ angular distribution
are the determination of the width difference $\Delta\Gamma_s$ from untagged 
data samples \cite{DDF} (for recent feasibility studies for the LHC, see 
\cite{belkov}), and the extraction of $\cos\delta_f\cos\phi_s$ terms, where 
the $\delta_f$ are CP-conserving strong phases. If we fix the signs of 
$\cos\delta_f$ through factorization, we may fix the sign of $\cos\phi_s$, 
which allows an {\it unambiguous} determination of $\phi_s$ \cite{DFN}. 
In this context, $B_s\to D_\pm\eta^{(')}$, $ D_\pm\phi$, ...\ decays are 
also interesting \cite{RF-gam-eff-03,RF-gam-det-03}.

\subsection{$B_s\to K^+K^-$}\label{ssec:BsKK}
The decay $B_s\to K^+K^-$ is dominated by QCD penguins and complements 
$B_d\to\pi^+\pi^-$ nicely, thereby allowing a determination of $\gamma$ 
with the help of $U$-spin flavour-symmetry arguments 
\cite{RF-BsKK}. Within the SM, we may write
\begin{equation}\label{fleischer-Bpipi-BsKK-ampl}
A(B_s^0\to K^+K^-)\propto\left[e^{i\gamma}+
\left(\frac{1-\lambda^2}{\lambda^2}\right)d'e^{i\theta'}
\right],
\end{equation}
where the hadronic parameter $d' e^{i\theta'}$ is the $B_s^0\to K^+K^-$
counterpart of the $B_d^0\to\pi^+\pi^-$ parameter $d e^{i\theta}$ 
introduced in (\ref{Bpipi-ampl}). In analogy to (\ref{ADIR-Bpipi})
and (\ref{AMIX-Bpipi}), we then have
\begin{eqnarray}
{\cal A}_{\rm CP}^{\rm dir}(B_s\to K^+K^-)&=&\mbox{fct}(d',\theta',\gamma)\\
{\cal A}_{\rm CP}^{\rm mix}(B_s\to K^+K^-)&=&\mbox{fct}(d',\theta',\gamma,
\phi_s).
\end{eqnarray}
As we saw above, $\phi_d$ and $\phi_s$ can 
``straightforwardly'' be fixed, also if NP should contribute to 
$B^0_q$--$\overline{B^0_q}$ mixing. Consequently, 
${\cal A}_{\rm CP}^{\rm dir}(B_d\to\pi^+\pi^-)$ and 
${\cal A}_{\rm CP}^{\rm mix}(B_d\to\pi^+\pi^-)$ allow us to eliminate
$\theta$, thereby yielding $d$ as a function of $\gamma$ in a 
{\it theoretically clean} way. Analogously, we may fix  $d'$ as a 
function of $\gamma$ with the help of 
${\cal A}_{\rm CP}^{\rm dir}(B_s\to K^+K^-)$ and 
${\cal A}_{\rm CP}^{\rm mix}(B_s\to K^+K^-)$. 

If we look at the corresponding Feynman diagrams, we observe that 
$B_d\to\pi^+\pi^-$ and $B_s\to K^+K^-$ are related to each other through 
an interchange of all down and strange quarks. Because of this feature, 
the $U$-spin flavour symmetry of strong interactions implies
\begin{equation}\label{fleischer-U-rel}
d=d', \quad \theta=\theta'.
\end{equation}
Applying the former relation, we may extract $\gamma$ and $d$ from the 
theoretically clean $\gamma$--$d$ and $\gamma$--$d'$ contours. 
Moreover, we may also determine $\theta$ and $\theta'$, allowing an 
interesting check of the second $U$-spin relation.
 
This strategy is very promising from an experimental point of view, since 
experimental accuracies for $\gamma$ of ${\cal O}(10^\circ)$ and 
${\cal O}(1^\circ)$ may be achieved at CDF-II and LHCb, respectively
\cite{TEV-BOOK,LHC-BOOK,LHCb}. As far as $U$-spin-breaking 
corrections are concerned, they enter the determination of $\gamma$ through 
a relative shift of the $\gamma$--$d$ and $\gamma$--$d'$ contours; their 
impact on the extracted value of $\gamma$ depends on the form of these 
curves, which is fixed through the measured observables. In the examples 
discussed in \cite{RF-Phys-Rep,RF-BsKK}, the result for $\gamma$ would be 
very robust under such corrections. For a more detailed discussion of
$U$-spin-breaking effects and recent attempts to estimate them, the
reader is referred to \cite{RF-BsKK,beneke,KMM-SU3}.

Interestingly, the quantity $H$ introduced in (\ref{H-def}) implies a very 
narrow SM ``target range'' in the 
${\cal A}_{\rm CP}^{\rm mix}(B_s\to K^+K^-)$--${\cal A}_{\rm CP}^{\rm dir}
(B_s\to K^+K^-)$ plane \cite{FlMa2}. A first important step to complement 
the analysis discussed in \ref{ssec:Bpipi} is the measurement of 
BR$(B_s\to K^+K^-)$, which is expected to be available soon from CDF-II. 
Once also the CP asymmetries of this channel have been measured, we may 
fully exploit the physics potential of the $B_s\to K^+K^-$, 
$B_d\to\pi^+\pi^-$ system as discussed above \cite{RF-BsKK}.

\subsection{$B_s\to D_s^{(\ast)\pm}K^\mp$}\label{sec-BsDsK}
Let us finally turn to colour-allowed ``tree'' decays of the kind 
$B_s\to D_s^{(\ast)\pm}K^\mp$, which have the interesting feature
that both a $B^0_s$ and a $\overline{B^0_s}$ meson may decay into
the same final state, thereby leading to interference between 
$B^0_s$--$\overline{B^0_s}$ mixing and decay processes, which involve 
the weak phase $\phi_s+\gamma$ \cite{BsDsK}. A similar feature is also 
exhibited by $B_d\to D^{(\ast)\pm}\pi^\mp$ modes, allowing us to probe 
$\phi_d+\gamma$ \cite{BdDpi}. Whereas the interference effects are
governed by $x_s e^{i\delta_s}\propto R_b\approx0.4$ in the $B_s$-meson
case and are hence favourably large, in the $B_d$ case they are described by 
$x_d e^{i\delta_d}\propto -\lambda^2R_b\approx-0.02$ and hence are tiny. 
These $B_s$ and $B_d$ modes can be treated on the 
same theoretical basis, and provide new strategies to determine $\gamma$ 
\cite{RF-gam-ca}. To this end, we may write them as 
$B_q\to D_q \overline{u}_q$,
where $D_s\in\{D_s^+, D_s^{\ast+}, ...\}$, $u_s\in\{K^+, K^{\ast+}, ...\}$
for $q=s$, and $D_d\in\{D^+, D^{\ast+}, ...\}$ , $u_d\in\{\pi^+, \rho^+, ...\}$
for $q=d$. We shall only consider $B_q\to D_q \overline{u}_q$ modes, 
where at least one of the $D_q$, $\overline{u}_q$ states is a pseudoscalar 
meson; otherwise a complicated angular analysis has to be performed.

In the ``conventional'' approach \cite{BsDsK,BdDpi}, the observables 
of the $\cos(\Delta M_qt)$ pieces of the time-dependent rate asymmetries
are used to extract the parameters $x_q$. To this end, ${\cal O}(x_q^2)$ 
terms have to be resolved. In the case of $q=s$, we have $x_s={\cal O}(R_b)$, 
implying $x_s^2={\cal O}(0.16)$, so that this may actually be possible, 
although challenging. On the other hand, $x_d={\cal O}(-\lambda^2R_b)$ 
is doubly Cabibbo-suppressed. Although it should be possible to resolve 
terms of ${\cal O}(x_d)$, this will be impossible for the vanishingly 
small $x_d^2={\cal O}(0.0004)$ terms, so that other approaches to fix 
$x_d$ are required \cite{BdDpi}. In order to extract $\phi_q+\gamma$, the 
mixing-induced observables provided by the $\sin(\Delta M_qt)$ terms
have to be measured. Following these lines, we arrive eventually
at an eightfold solution for $\phi_q+\gamma$. If we fix the
sign of $\cos\delta_q$ with the help of factorization, a fourfold 
discrete ambiguity emerges. In particular, we may also extract the
sign of $\sin(\phi_q+\gamma)$, which allows us to distinguish between the 
two solutions in Fig.~\ref{fig:rho-eta-Bpipi}. In these considerations, the 
angular momentum $L$ of the $D_q\overline{u}_q$ state has to be properly 
taken into account \cite{RF-gam-ca}.

Let us now discuss other new features of the $B_q\to D_q \overline{u}_q$ 
modes, following \cite{RF-gam-ca}. As we noted above, $\Delta\Gamma_s$ 
may provide interesting ``untagged'' observables. If we combine them with 
the ``tagged'' mixing-induced observables provided by the $\sin(\Delta M_st)$ 
terms, we may extract, in a simple manner, $\tan(\phi_s+\gamma)$, which 
gives an {\it unambiguous} value for $\phi_s+\gamma$ itself if we fix 
again the sign of $\cos\delta_s$ through factorization. Another important 
advantage of this new strategy is that only observables proportional to 
${\cal O}(x_s)$ are employed, i.e.\ no $x_s^2$ terms have to be resolved. 
Another interesting feature of the $B_q\to D_q \overline{u}_q$
system is that we may obtain bounds on $\phi_q+\gamma$, which may be
highly complementary for the $B_s$ and $B_d$ modes, thereby implying
particularly narrow, theoretically clean ranges for $\gamma$. Whereas the 
$B_s$ decays are not yet accessible, first results for the
$B_d\to D^{(\ast)\pm}\pi^\mp$ modes obtained by BaBar give
$|\sin(\phi_d+\gamma)|>0.87$ (68\% C.L.) and 
$|\sin(\phi_d+\gamma)|>0.58$ (95\% C.L.) \cite{BaBar-BDpi}. The analysis
of these channels at Belle is also in progress \cite{Belle-BDpi}.
If we look at the $B_s^0\to D_s^{(\ast)+}K^-$ and 
$B_d^0\to D^{(\ast)+}\pi^-$ decay topologies, we observe that they are 
related to each other through an interchange of all down and strange quarks.
Consequently, the $U$-spin flavour symmetry of strong interactions implies
relations between the corresponding hadronic parameters, which can be 
implemented in a variety of ways. Apart from features related to multiple 
discrete ambiguities, the most important advantage of this strategy with 
respect to the ``conventional'' approach is that the experimental resolution 
of the $x_q^2$ terms is not required. In particular, $x_d$ does {\it not} 
have to be fixed, and $x_s$ may only enter through a $1+x_s^2$ correction, 
which can straightforwardly be determined through untagged $B_s$ rate 
measurements. In the most refined implementation of this strategy, the 
measurement of $x_d/x_s$ would {\it only} be interesting for the inclusion 
of $U$-spin-breaking effects. Moreover, we may obtain interesting insights 
into hadron dynamics and $U$-spin-breaking effects. 

In order to explore CP violation, the colour-suppressed counterparts
of the $B_q\to D_q \overline{u}_q$ modes are also very interesting. 
In the case of the $B_d\to D K_{\rm S(L)}$, $B_s\to D \eta^{(')}, D \phi$, 
...\ modes, we may extract $\tan\gamma$ in an elegant and {\it 
unambiguous} manner, whereas $B_s\to D_\pm K_{\rm S(L)}$, 
$B_d\to D_\pm \pi^0, D_\pm \rho^0$, ...\ modes allow very interesting 
determinations of $\phi_q$ with theoretical 
accuracies one order of magnitude higher than those of
the conventional  $B_d\to J/\psi K_{\rm S}$, $B_s\to J/\psi \phi$
approaches. In particular, $\phi_s^{\rm SM}=-2\lambda^2\eta$ could be 
determined with only ${\cal O}(1\%)$ uncertainty 
\cite{RF-gam-eff-03,RF-gam-det-03}.

\section{Conclusions and Outlook}\label{sec:concl}
Thanks to the $B$ factories, CP violation is now a well established 
phenomenon in the $B$ system, and many strategies to explore 
CP violation can be confronted with the data. Although the measurement 
of $\sin\phi_d$ through $B_d\to J/\psi K_{\rm S}$ agrees with the SM -- but 
leaves a twofold solution for $\phi_d$ itself -- the current $B$-factory
data point towards certain ``puzzles'', for instance in 
$B\to\pi K$ and $B_d\to \phi K_{\rm S}$ decays. It will 
be very exciting to see whether these potential discrepancies with 
the SM will survive improved measurements. Another important aspect 
of the exploration of CP violation is the $B_s$-meson system, which will 
be accessible at run II of the Tevatron and can be fully exploited in 
the era of the LHC. Certainly a promising future of CP violation and
quark-flavour physics is ahead of us!

\section*{Acknowledgements}
I would like to thank the organizers, in particular Giuseppe Nardulli
and Pietro Colangelo, for inviting me to this very interesting 
and extremely pleasant workshop.

\end{document}